\documentclass[twocolumn]{cpcauth}

\usepackage{times}

\usepackage{graphicx}

\usepackage{amssymb}
\newcommand{\dir}{Figs}

\begin{document}

\begin{frontmatter}


\title{Fluctuations and Defects in Lamellar Stacks of Amphiphilic Bilayers}

\author[ia1,ia2,ia3]{Claire Loison}
\author[ia3]{Michel Mareschal}
\author[ia2]{Friederike Schmid\corauthref{cor1}}
\corauth[cor1]{Corresponding Author:}
\ead{schmid@physik.uni-bielefeld.de}
\address[ia1]{MPI f\"ur Chemische Physik fester Stoffe, 
N\"othnitzer Str. 40, D-01187 Dresden }
\address[ia2]{Fakult\"at f\"ur Physik, Universit\"at Bielefeld,
Universit\"atsstra\ss e 25, D-33615 Bielefeld}
\address[ia3]{Centre Europ\'een de Calcul Atomique et Mol\'eculaire,
ENS Lyon, 46, All\'ee d'Italie, F-69007 Lyon}

\begin{abstract}
We review recent molecular dynamics simulations of thermally activated
undulations and defects in the lamellar $L_\alpha$ phase of a 
binary amphiphile-solvent mixture, using an idealized molecular 
coarse-grained model: Solvent particles are represented by beads, 
and amphiphiles by bead-and-spring tetramers. We find that our results 
are in excellent agreement with the predictions of simple mesoscopic 
theories: An effective interface model for the undulations, and a line 
tension model for the (pore) defects. We calculate the binding rigidity
and the compressibility modulus of the lamellar stack as well as the
line tension of the pore rim. Finally, we discuss implications 
for polymer-membrane systems.
\end{abstract}

\end{frontmatter}

\section{Introduction}
\label{sec1} 
Amphiphilic molecules, such as lipids, contain hydrophilic 
(water loving) and hydrophobic (water fearing) parts. 
In water, they self-assemble spontaneously such that the 
hydrophobic parts are shielded from the water environment. 
One of the most prominent structures is the lamellar phase, 
where the amphiphiles are arranged in stacks of bilayers. 
Lipid bilayers are of special interest, because they are essential 
components of biological membranes. 

Phenomenologically, membranes are often described as
two dimensional surfaces. A stack of nearly planar membranes 
of average distance $\bar{d}$ can be parametrized by a 
set of functions $u_n ({\bf x})$, where ${\bf x}=(x_1,x_2)$ 
are coordinates in the plane and $x_3=u_n + n \bar{d}$ denotes
the position of the $n$th membrane in the direction 
perpendicular to the plane. The simplest theoretical
Ansatz approximates the free energy of such a stack 
by~\cite{Lei_JP_95}
\begin{equation}
\label{Fd}
E_{\mbox{el}} =   \sum_n \!\! \int_A \!\! \mbox{d}^2x 
\left\{
\frac{K_c}{2} ( \Delta u_n )^2 + 
 \frac{B}{2} (  u_{n+1}\!\! - \!\! u_{n} )^2
\right\}
\end{equation}
with the bending rigidity $K_c$ and the compressibility $B$.
This free energy functional describes thermally activated 
membrane undulations with a characteristic in-plane 
correlation length $\xi = (K_c/B)^{1/4}$.

Apart from undulations, membrane stacks also contain
defects. Among these, the membrane pores are particularly 
important, because they influence crucially the permeation 
properties of the membrane. In the simplest 
phenomenological model for pore formation~\cite{Lister_PL_75}, 
a pore with the area $a$ and the contour length $c$ has 
the free energy
\begin{equation}
\label{energy_pore}
E_{\mbox{pore}}=E_0+\lambda c -\gamma a,
\end{equation}
where $\gamma$ is the surface tension of the membrane and 
$\lambda$ a line tension.  In an isotropic membrane stack, 
the membranes are tensionless with $\gamma = 0$.

In the present paper, we describe large scale molecular
dynamics simulations of a molecular coarse-grain model
which allow to test these simple theoretical concepts.
We find that the models models (\ref{Fd}) and 
(\ref{energy_pore}) capture the physics of our membrane
stacks very well. The results are then used to discuss 
a more complex system, {\em i.e.}, a polymer inserted
in a membrane stack.

\section{Simulation Model, Method, and Data Analysis}
\label{sec2} 

The simulation model was derived from a similar model
by Soddemann et al~\cite{Soddemann_EPJE_01}. The ``molecules'' consist 
of two types of beads, $H$ or $P$. ``Solvent'' particles are
single $P$ beads, ``Amphiphiles'' are $H_2 P_2$ tetramers.
All beads have the same mass $m$ and diameter $\sigma$ 
and interact with a repulsive hard core potential. 
In addition, beads of the same type attract each other. 
At appropriate parameter values, the amphiphiles 
self-assemble spontaneously and form a lamellar
stack of bilayers.

Our simulations were carried out in the NPT 
ensemble with a Langevin thermostat. All sides of the simulation 
box were allowed to fluctuate in order to ensure isotropic
pressure. We studied systems containing a total number of 
153600 beads, 30720 of which were solvent particles,
at pressure $P = 2.9 k_B T/\sigma^3$.  
The total run length was $10^7$ MD steps, 
corresponding to the time span $10^5 \sqrt{m \sigma^2/k_B T}$.
The initial equilibration time was $10^6$ MD steps.
Further details can be found in Refs.~\cite{Loison_1,Loison_2}.

%

To analyze the data, the simulation box was subdivided
into columns of diameter $\sim 1.3 \sigma$, perpendicular to 
the membranes. For each configuration and in each column, 
the local profiles were determined. This allowed to localize the 
membranes and to identify membrane defects~\cite{Loison_1,Loison_2}.

\section{Results}
\label{sec3} 

\noindent
{\em Membrane Structure and Thermal Fluctuations}

\bigskip

\begin{figure}[tbp]
\includegraphics[scale=0.27,angle=-90]{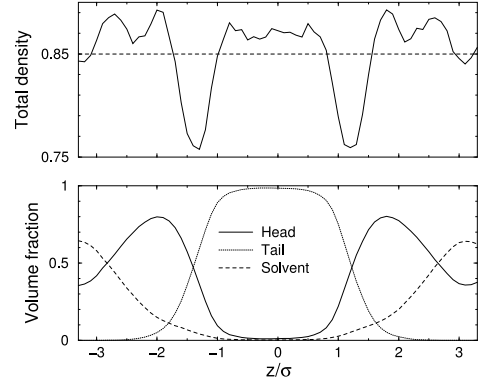}
\caption{\label{fig:profiles} 
Total density profiles (top) and partial volume fractions
of head, tail and solvent beads. (Head and tail beads
refer to $P$ and $H$ beads of amphiphiles).
}
\end{figure}

The local membrane structure can be characterized by 
the local density and composition profiles. In order to 
eliminate the smearing effect of the
undulations, we calculate the profiles in each column 
separately and shift it by the local interface position before 
averaging. The results are shown in Fig. \ref{fig:profiles}. 
The total density varies very little throughout the system, 
but the amphiphile and solvent layers are nevertheless
well-separated. 

\begin{figure}[bp]
\includegraphics[scale=0.2,angle=-90]{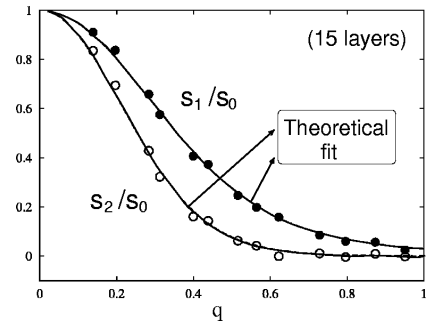}
\caption{\label{fig:struc} Transmembrane structure factor,
fitted to the prediction of the model \ref{Fd} with one
fit parameter $\xi = (K_c/B)^{1/4}$.
}
\end{figure}

From the statistics of the interface positions, we can
analyze the capillary wave spectrum. Here we show only
the transmembrane structure factor between membranes,
which we define as 
$
S_n(q) = \langle u_n({\bf q})^* u_0({\bf q}) \rangle,
$
where $u_n({\bf q})$ is the two dimensional Fourier transform
of $u_n({\bf x})$. Within the model (\ref{Fd}), 
we can calculate the quantity $S_n(q)$ theoretically and 
compare the results with the simulation data. The only fit 
parameter is the parallel correlation length 
$\xi = (K_c/B)^{(1/4)}$. Fig. \ref{fig:struc} shows that 
the theoretical prediction fits the data very well at 
$\xi = 2.34 \sigma$.

Similar fits allow to determine the layer distance 
$\bar{d} = 6.38 \sigma $, and the Caill\'e parameter 
$\eta_1 = \pi k_B T/2 \bar{d}^2 \sqrt{K_c B} = 0.053$.
This gives us all the parameters of the model (\ref{Fd}),
{\em i.e.}, $K_c = 4 k_B T$ and $B = 0.13 k_B T/\sigma^4$.

\bigskip

\noindent
{\em Pore Defects}

\bigskip

The main defects in our system were pores, and only these 
were studied in detail. The density profiles of the pores 
(not shown here, see Ref.~\cite{Loison_2}) show unambiguously 
that larger pores are hydrophilic, i.e., the amphiphiles 
at the pore rim rearrange themselves such that the solvent 
in the pore is shielded from the hydrophobic molecules tails. 
Since our amphiphiles are very short, the energy penalty
on pores is low and the number of pores in the membranes is
comparatively large. Nevertheless, we have established
in several tests that the pores are essentially
uncorrelated. Therefore, pore interactions can be
neglected, and we can attempt to interpret our data
in terms of the model (\ref{energy_pore}) with $\gamma=0$.

\begin{figure}[tbp]
\includegraphics[scale=0.23,angle=-90]{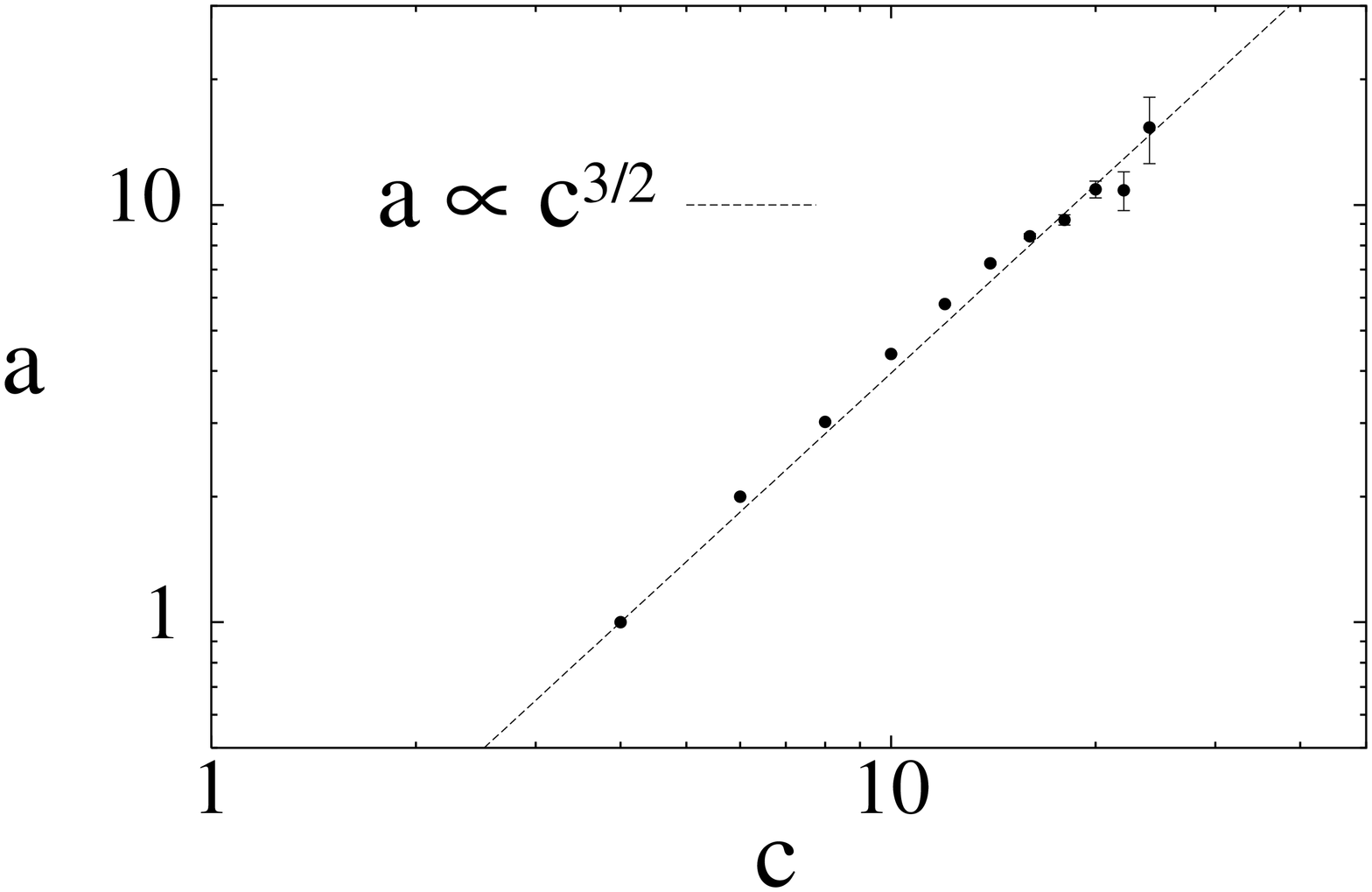}\\
\includegraphics[scale=0.28,angle=-90]{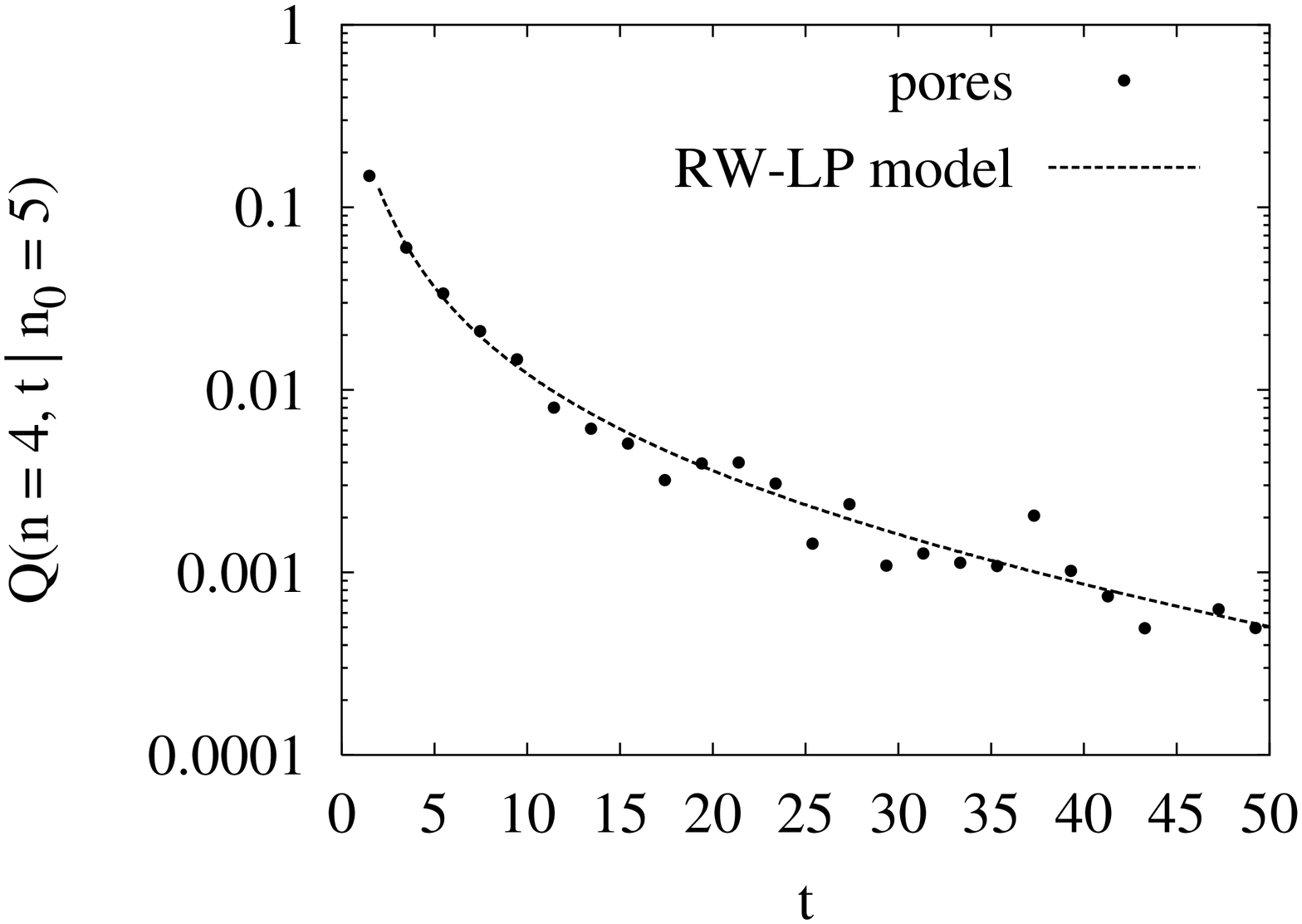}
\caption{\label{fig:pores} 
Tests of the line tension model (\ref{energy_pore}).
Top: Mean area $a$ of pores as a function of contour
length $c$.
Bottom:
Probability that a pore with contour length $c=10$ 
reaches the contour $c=8$ for the first time
at the time $t$, compared with the theoretical 
prediction for a random walk in a linear 
potential (RW-LP model~\cite{Khanta_Pramana_83}).
}
\end{figure}

The validity of Eq.~\ref{energy_pore} was tested in 
several ways. First, we verified that the pore contour 
lengths $c$ were distributed according to a Boltzmann 
distribution. This allowed to extract the value of 
the line tension, $\lambda = 0.7 k_B T/\sigma$. 
Next we studied the shape of the pores. Since the membranes 
have no surface tension, the pores are not circular: 
At $\gamma=0$, Eq.~\ref{energy_pore} describes an ensemble 
of self avoiding ring chains.
These should have by the fractal dimension of two-dimensional
self-avoiding polymers, characterized by the Flory
exponent $\nu_2=3/4$. For example, the area $a$ of the 
pores scales with the contour length like 
$a \propto c^{3/2}$. Fig.~\ref{fig:pores} a) shows
that the simulation data are compatible with this
scaling behavior.

The energy model \ref{energy_pore} also has implications
for the pore dynamics: The dynamical evolution of the 
pore contour lengths should correspond to a random walk 
in a linear potential, $E = E_0 + \lambda c$. This 
stochastic model has been solved analytically~\cite{Khanta_Pramana_83}
and we can compare our data with exact results, {\em e.g.},
for first passage time distributions.
As Fig.~\ref{fig:pores} b) demonstrates, the agreement
is again very satisfactory.

\section{Outlook: Polymer-Membrane Systems }

\begin{figure}[tbp]
\includegraphics[scale=0.22]{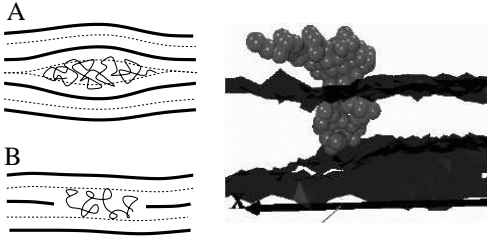}
\caption{ \label{fig:polymer} 
Left:
Possible ways of incorporating a polymer in
a membrane stack: 
A -- Squeezed between two membranes;
B -- Spanning several layers and creating a hole. 
Right:
Configuration snapshot of a polymer of length
$N=100$ in the membrane stack. The surfaces show 
the mid-planes of two membranes close to the polymer. 
}
\end{figure}

The success of the phenomenological description
(\ref{Fd}) and (\ref{energy_pore}) motivates 
its application to more complex situations. 
As an example, we shall now discuss the effect of 
inserting a hydrophilic polymer in our system. 
Two possible scenarios are illustrated in 
Fig.~\ref{fig:polymer} (left): Either the polymer stays 
squeezed between two membranes and deforms them locally, 
or it spans between layers, creating one or several holes. 

We have performed simulations of a membrane stack 
containing a polymer with $100$ $H$-beads~\cite{tobepublished}.
In free $H$-solution, the polymer exhibits
self-avoiding walk statistics with the Kuhn length 
$a=0.4 \sigma$. A configuration in a membrane stack
is shown in Fig.~\ref{fig:polymer} (right). 
The polymer induces a hole in one of the adjacent
membranes and collapses into a globule which 
spreads between two layers.

The situation can be analyzed using scaling
arguments. The entropy penalty for confining
a polymer of length $N$ into a disk of width
$R_{\perp}$ and diameter $R_{\parallel}$ 
is~\cite{Daoud_JP_77,tobepublished}
\begin{eqnarray}
\label{poly_2d}
E_{\mbox{c}} &\sim&
N ({a}/{R_{\perp}})^{5/3} +
N^3 {a^5}/({R_{\perp} R_{\parallel}^4})
\quad 
R_{\parallel} > R_{\parallel}^c \\
\label{poly_3d}
E_{\mbox{c}} &\sim&
N^{9/4} ({a^3}/{R_{\perp} R_{\parallel}^2})^{5/4} 
\qquad \qquad  \quad R_{\parallel} < R_{\parallel}^c 
\end{eqnarray}
with $R_{\parallel}^c \sim N^{1/2} R_{\perp}^{1/6} a^{5/6}$.
Eq.~\ref{poly_2d} corresponds to a 2D compressed (pancake) regime,
where the main entropic penalty comes from the confinement 
in the $R_{\perp}$ direction (first term),
and the polymer behaves like a slightly compressed
2D self avoiding walk in the $R_{\parallel}$ direction
(second term). Eq.~\ref{poly_3d} describes a 3D compressed 
regime. 

In the scenario A of Fig.~\ref{fig:polymer},
$E_{\mbox{p}}$ has to be balanced with the 
elastic energy $E_{\mbox{el}}$ of the membrane stack.  
From Eq.~\ref{Fd}, we can derive~\cite{tobepublished}
$E_{\mbox{el}} \sim \pi K_c (R_{\parallel} \: \delta)^2/\xi^4$,
where $\delta$ is the deformation of the membranes closest 
to the polymer. 
In the scenario B, the elastic energy can be neglected, 
and the polymer compression energy has to be balanced with 
the energy of pore formation, 
$E_{\mbox{pore}} \sim 2 \pi \lambda R_{\parallel}$.

Details of the calculation will be given 
elsewhere~\cite{tobepublished}. Here we merely sketch 
some of the results: Short chains remain squeezed between 
membranes (scenario A) and assume a 2D compressed (pancake)
structure. In this regime, the total energy scales 
linearly with $N$.  Long chains collapse into a 3D compressed 
globule, which creates one pore and spans exactly two layers 
(scenario B).  Here, the total energy scales with $N^{9/14}$. 
In stacks with low compressibility (large $B \bar{d}^4$), 
2D compressed states spanning two or several layers may appear
at intermediate chain lengths. 

This qualitatively explains the observation in the
simulation: The polymer is in the long chain regime, 
collapses into the 3D compressed state and spreads 
between two layers. The agreement is not yet quantitative.
According to the scaling theory, the transition to the 
long-chain regime is expected at the chain length 
$N_c \sim \lambda^2 a^{-5/3} \bar{d}^{11/3}$, which 
corresponds to $N \sim 2000$ with our parameter
values. Our chain is much shorter. The discrepancy 
can presumably be explained by the unknown prefactor 
in the scaling law for $N_c$. 

To conclude, the phenomenological models (\ref{Fd}) and
(\ref{energy_pore}) are useful to analyze not only pure
membrane stacks, but also polymer-membrane systems. 
It is remarkable that both the scaling theory and the 
simulations predict that long (non-adsorbing) hydrophilic 
polymers enforce pore formation in the membranes.\\ 

\renewcommand{\baselinestretch}{0.9}\small

We thank Ralf Eichhorn, Ralf Everaers, Kurt Kremer, 
Peter Reimann, Thomas Soddeman, and Hong-Xia Guo for 
stimulating discussions. The simulations were carried 
out at the computer centers of the Max-Planck
society (Garching) and the Commissariat \`a 
l'Energie Atomique (Grenoble).

\vspace*{-0.5cm}

\renewcommand{\baselinestretch}{1.}\normalsize

\bibliography{paper}

\end{document}